\documentclass{osa-article}

\journal{oe}

\usepackage{siunitx}
\DeclareSIUnit\wn{\raiseto{-1}\cm}
\DeclareSIUnit\mum{\micro\m}
\DeclareSIUnit{\sqrthz}{\ensuremath{\sqrt{\text{\hertz}}}}
\newcommand{\ur}[1]{\text{#1}}
\newcommand{\vnu}{\tilde{\nu}}
\articletype{Research Article}

\begin{document}

\title{Nonlinear interferometer for Fourier-transform mid-infrared gas spectroscopy using near-infrared detection}

\author{Chiara Lindner,\authormark{*} Jachin Kunz, Simon J. Herr, Sebastian Wolf, Jens Kie\ss ling and Frank K\"{u}hnemann}

\address{Fraunhofer Institute for Physical Measurement Techniques IPM, Georges-K\"{o}hler-Allee 301, 79110 Freiburg, Germany}

\email{\authormark{*}chiara.lindner@ipm.fraunhofer.de} 



\begin{abstract}
Nonlinear interferometers allow for mid-infrared spectroscopy with near-infrared detection using correlated photons. Previous implementations have demonstrated a spectral resolution limited by spectrally selective detection. In our work, we demonstrate mid-infrared transmission spectroscopy in a nonlinear interferometer using single-pixel near-infrared detection and Fourier-transform analysis. A sub-wavenumber spectral resolution allows for rotational-line-resolving spectroscopy of gaseous samples in a spectral bandwidth of over 700\,cm$^{-1}$. We use methane transmission spectra around \SI{3.3}{\um} wavelength to characterize the spectral resolution, noise limitations and transmission accuracy of our device. The combination of nonlinear interferometry and Fourier-transform analysis paves the way towards performant and efficient mid-infrared spectroscopy with near-infrared detection.
\end{abstract}

\section{Introduction}
Nonlinear interferometers offer new perspectives in metrology, relying on the quantum effects observed in the interference phenomena of correlated photons. A typical source of correlated photons is spontaneous parametric down-conversion (SPDC) inside a nonlinear-optical crystal. This effect can be described as the spontaneous decay of pump photons inside a nonlinear medium into two photons, called signal and idler. The sum of the energies of signal and idler photons equals the energy of the pump photon.
If the pump, signal and idler photons of one SPDC source pass through a second, identical nonlinear crystal, interference can be observed between the two indistinguishable correlated photon sources for both signal and idler photons.
Due to induced coherence \cite{Zou.1991}, the visibility and phase of the interference intensity pattern of the signal photons depend on the transmission and phase of all three beams: pump, signal and idler \cite{Burlakov.1997, Chekhova.2016, Miller.2019, Lavoie.2020}.\\
Recently, nonlinear interferometers using a large spectral separation between signal and idler photons have sparked interest, being able to separate sample interaction and light detection into different spectral regions. Nonlinear interferometers have been demonstrated for applications such as imaging \cite{Lemos.2014}, sensing \cite{Kutas.2020}, microscopy \cite{Kviatkosvky.2020, Paterova.2020}, ellipsometry \cite{Paterova.2019} and optical coherence tomography \cite{Paterova.2018, Vanselow.2019, Machado.2020}.\\
Mid-infrared spectroscopy is an important analytic technique with a wide range of applications. Novel concepts such as spectroscopy by nonlinear optical frequency upconversion \cite{Tidemand.2016,Friis.2019,Wolf.2017} or frequency-comb-based spectroscopy \cite{Picque.2019, Ycas.2018, Villares.2014} seek to push beyond limitations of conventional broadband infrared spectroscopic techniques such as Fourier-transform spectroscopy (FTIR).
In this context, nonlinear interferometers using pairs of correlated mid-infrared and near-infrared/visible photons may offer an alternative approach for new measurement concepts \cite{Kalashnikov.2016b, Paterova.2017, Paterova.2018b, Lindner.2020}, as they combine the generation of the infrared photons for sample interaction with the short-wavelength photons for detection.
The use of silicon-based visible or near-infrared detectors promises lower dark noise and higher bandwidth in comparison to infrared detectors, which often require cooling.\\
One key element of spectroscopic nonlinear interferometers is the retrieval of the infrared spectral information via detection of the visible or near-infrared signal interference. 
Using a monochromator for spectral selective detection of the visible signal light, a spectral resolution of \SI{5.2}{\wn} was demonstrated for idler wavelengths around \SI{2200}{\nm} \cite{Paterova.2018b}.
In a recent publication, it was demonstrated that the infrared spectral information can also be obtained from a Fourier-transform of the detected signal interference pattern with a comparable spectral resolution of \SI{6}{\wn} \cite{Lindner.2020}. To this purpose, a series of signal interference patterns was recorded using a silicon-based camera while varying the delay between the signal and idler interferometer arm. Hereby, a bandwidth of \SI{100}{\wn} was realized by using non-collinear SPDC emission in periodically-poled lithium niobate (PPLN). This approach required spatially resolved detection of the signal interference pattern with a camera and pixel-wise analysis due to a spatially varying phase and optical path difference for the non-collinear components.\\
In this work, we overcome these limitations by using an SPDC source designed for broadband collinear emission (cf. \cite{Vanselow.2019b, Kviatkosvky.2020}). This allows using a single-pixel (near-infrared) silicon-based detector to accurately analyze the mid-infrared transmission of a sample using a simple Fourier-transform analysis.
In consequence, the possible spectral resolution is only limited by the maximum interferometer delay, opening the door for sub-wavenumber resolution, which enables spectroscopic analysis of gaseous samples resolving individual rotational absorption lines. In contrast to previous works, a spectral range of \SIrange{3.1}{4.0}{\um} wavelength (\SI{725}{\wn}) is covered in a single measurement with a spectral resolution of up to \SI{0.56}{\wn}. We demonstrate this approach with transmission spectra of a dilution of methane (CH$_4$) in nitrogen (N$_2$), which are used for a performance characterization of the presented setup. The spectroscopic accuracy is evaluated in a quantitative comparison to a simulation based on the HITRAN spectroscopic database.

\section{Experimental realization}

\begin{figure}[t]
	\centering
	\fbox{\includegraphics[scale=0.9]{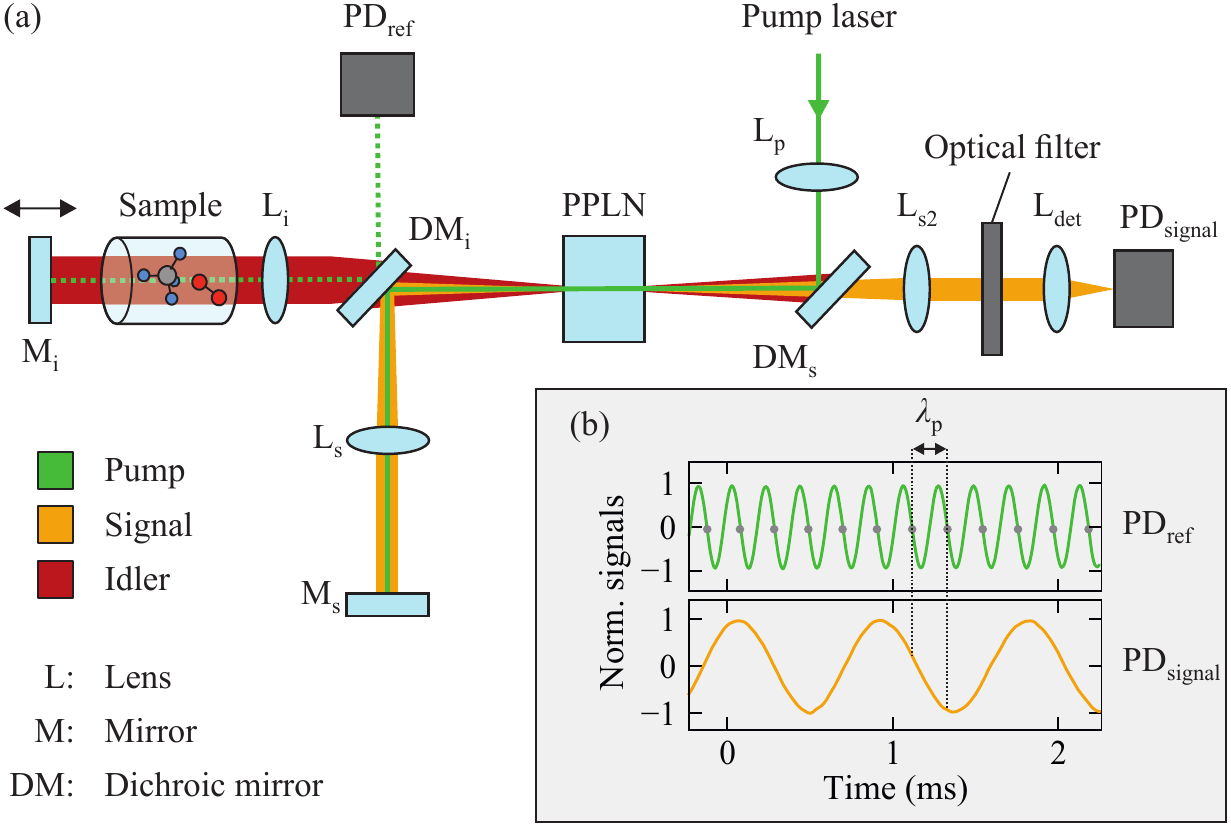}} 
	\caption{(a) Setup of the nonlinear Michelson-interferometer. (b) Zoom into measured detector signals: reference pump interference (green) and signal interference (orange), modulated by movement of the idler mirror M$_{\ur{i}}$. Every second zero-crossing (grey dots) of the pump interference signal is used as a reference for an equidistantly ($\lambda_{\ur{p}}$) sampled interferogram.}
	\label{fig:setup}
\end{figure}

As a source of correlated photons we use SPDC in a 10-$\ur{mm}$-long PPLN crystal with 5-mol\%-MgO doping.
Broadband collinear SPDC emission is achieved using signal and idler wavelengths with matching group indices, as described in Ref. \cite{Vanselow.2019b}. 
MgO-doped lithium niobate has the same group index for wavelengths around \SI{3.6}{\mum} (idler) and \SI{1}{\mum} (signal).
From energy conservation, the pump wavelength results to \SI{785}{\nm}, phase matching is achieved using a poling period of \SI{21.5}{\mum} at \SI{65}{\degreeCelsius} crystal temperature. Using this phase matching configuration, an idler wavelength range of \SIrange{3.1}{4.0}{\um} is emitted with correlated signal emission below \SI{1.1}{\um} wavelength, so that silicon-based detectors with low dark noise can be used.\\
A schematic setup of the nonlinear interferometer is shown in Fig. \ref{fig:setup}. All optical lenses (except $\ur{L}_\ur{i}$) are biconvex glass (NBK-7) lenses and anti-reflection coated for pump and signal wavelengths.
As a pump source, we use a tunable continuous-wave Ti:sapphire laser (\textit{M Squared SolsTiS}) with an output power of \SI{700}{\mW}, which is frequency stabilized to \SI{785}{\nm} wavelength (with an accuracy of \SI{10}{\MHz}) using an external wavelength meter. The pump laser beam passes an optical isolator and is focused with lens $\ur{L}_\ur{p}$ to a beam diameter of about \SI{70}{\um} at the center of the nonlinear crystal (PPLN). The pump beam is reflected by the dichroic mirror DM$_\ur{s}$ \textit{(Thorlabs DMLP900)} and illuminates the nonlinear crystal. The crystal is stabilized to a temperature of \SI{65}{\degreeCelsius} using a Peltier element. The signal and idler photons created by SPDC are then separated by the dichroic mirror DM$_\ur{i}$ (YAG-substrate with custom coating).\\ 
The mid-infrared idler beam is transmitted by DM$_\ur{i}$ and then collimated with a CaF$_2$ lens (L$_\ur{i}$) with a focal length of \SI{100}{\mm}. The plane gold mirror M$_\ur{i}$ placed in \SI{100}{\mm} distance to L$_\ur{i}$ reflects the idler light, which is imaged back into the center of the nonlinear crystal in an effective 4$f$ relay optic. The mirror M$_\ur{i}$ is mounted on a voice-coil translation stage (\textit{PI V-524.1}), which allows for a maximum displacement of $\pm$\SI{10}{\mm} in beam direction with a position accuracy (repeatability) of \SI{20}{\nm}. The interferometer is adjusted so that the point of zero delay (ZDP) between the interferometer arms is at zero displacement of M$_\ur{i}$.\\
The signal and pump beams, which are reflected by the dichroic mirror DM$_\ur{i}$, take an analogue path in the second interferometer arm, which has a fixed 4$f$ optical path length. The signal and pump light is collimated with lens L$_\ur{s}$ with \SI{100}{\mm} focal length and reflected by the plane dielectric mirror M$_\ur{s}$, which images the signal light back into the nonlinear crystal. The back-reflected pump laser causes a second SPDC process, leading to interference of the signal and idler photons generated by either pass through the crystal. 
Using a silicon-based photodiode power sensor (\textit{Thorlabs S130C}), a total signal SPDC power of \SI[separate-uncertainty=true]{20,0(14)}{\nano\W} (double pass, same setting as for the spectroscopic measurements) was measured from which we can estimate a corresponding infrared idler power of about \SI{6}{\nano\W}. It is to be noted that blocking the idler beam inside the nonlinear interferometer has no measurable effect on the total signal power of the second SPDC process (cf. \cite{Zou.1991,Lemos.2014}), which verifies that the two SPDC processes are indeed spontaneous and without any stimulated emission, due to low parametric gain. For a nonlinear interferometer in the low gain regime, the SPDC power depends linearly on the available pump power and the interference contrast depends linearly on the transmission of all three beams (pump, signal and idler) \cite{Chekhova.2016}.\\
After the second pass, the pump beam is again reflected by DM$_\ur{s}$ and removed by the optical isolator. The idler light is absorbed by the dichroic mirror. The overlapping signal light is transmitted by the dichroic mirror and collimated by lens L$_\ur{s2}$ with \SI{100}{\mm} focal length. Residual pump and ambient light is removed by an optical \SI{850}{\nm} long pass filter. The lens L$_\ur{det}$ focuses the signal light onto the active area of the detector PD$_{\ur{signal}}$. As a detector we use a silicon avalanche photodiode (APD, \textit{Laser Components A-CUBE-S500-01}) with a bandwidth of \SI{1}{\MHz} and a noise equivalent power of \SI{6}{\femto\W/\sqrthz} (specified at \SI{905}{\nm} detection wavelength). At a gain factor of $M=100$, the responsivity is specified to \SI{45}{\A/\W} at \SI{1}{\um} wavelength, which corresponds to a quantum efficiency $\epsilon$ of $55.8\,\%$.
For noise reduction, the detector signal passes an electronic filter with \SIrange{0.4}{5}{\kHz} pass band. The signal is then digitized with an analog-digital converter.\\ 
A maximum interference contrast of 15\,\% is measured as the peak AC/DC ratio of the detector signal without the electronic bandpass filter. 
According to simulations the achievable contrast is limited to about 20\,\% by dispersion inside the nonlinear crystal; 
additional losses are expected due to the non-ideal reflective and anti-reflective coatings of the optical elements and the imperfect beam overlap of the two SPDC processes.\\
The sample placed in the idler beam is a gas cell which can be filled with an analyte or pure nitrogen as reference. The cell consists of a small cylinder with \SI{15}{\mm} free aperture and \SI{20}{\mm} of interaction length. The cell windows consist of BaF$_2$ and are anti-reflective coated for the idler wavelength range.\\
For obtaining the spectrum of the interferometer with a Fourier-transform, the interferogram has to be sampled at equidistant positions of M$_\ur{i}$. Setups of classical Fourier-transform infrared (FTIR) spectrometers often use an additional HeNe-laser as reference, since the sinusodial interferogram of a near-monochromatic source has its zero-crossings at intervals which correspond to a retardation of half its wavelength \cite{Griffiths.2007}. In our setup, we take advantage of the residual transmission of the pump beam through the dichroic mirror DM$_\ur{i}$, which is shown as a dashed line in Fig. \ref{fig:setup}(a). With movement of mirror M$_\ur{i}$ the small portion of pump light (<1\,\%) receives the same retardation as the idler light (neglecting the dispersion in air). When the weak pump light is then reflected on DM$_\ur{i}$, it is superimposed with the residual transmission of the pump light which was originally reflected on the dichroic mirror. Thus the dichroic mirror DM$_\ur{i}$ and the end mirrors M$_\ur{s}$ and M$_\ur{i}$ form a classical Michelson interferometer for the pump beam, with a high amplitude contrast between the interferometer arms. The weak pump beam interference pattern is measured with a photodiode PD$_\ur{ref}$. The electronic signal is passed through a high-pass filter in order to remove the DC component and then digitized. The inset in Fig. \ref{fig:setup}(b) shows a zoom into recorded measurement signals, referenced to their acquisition time stamp. The nonlinear interference detector signal (orange curve) is evaluated at every second zero crossing (gray dots) of the reference signal (green curve) to obtain the required equally-spaced data points as input for the Fourier-transformation.

\section{Measurement and analysis procedure}
\begin{figure}[t!]
	\centering
	\fbox{\includegraphics{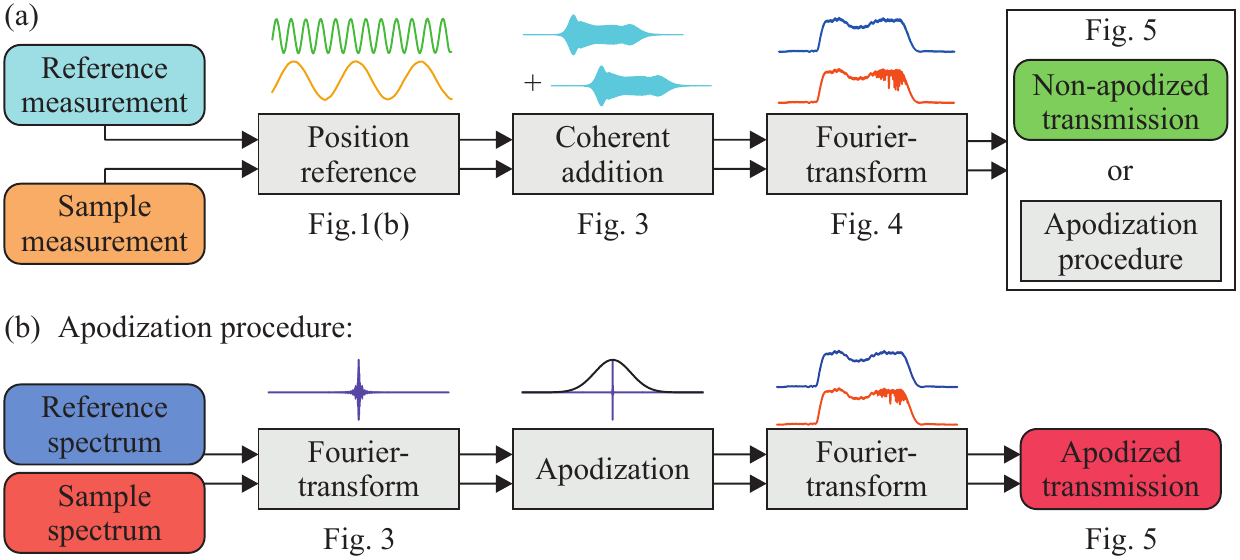}}
	\caption{Overview over the analysis procedure: (a) Steps in the treatment of the interferograms of reference and sample measurement. From the modulus (power spectral density) of the Fourier-transformed spectra, the non-apodized transmission spectrum can be calculated directly; alternatively, spectra using apodization can be calculated with the steps shown in (b).}\label{fig:analysis}
\end{figure}

\begin{figure}[bt]
	\centering
	\fbox{\includegraphics[scale=0.9]{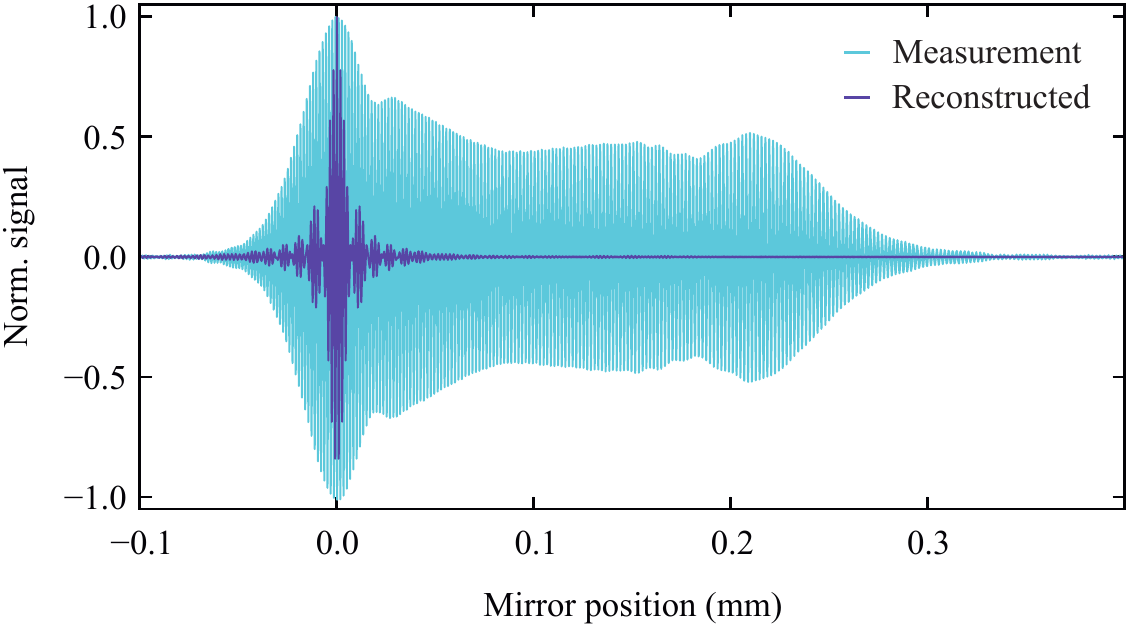}} 
	\caption{Averaged interferogram of 100 reference measurement scans (light blue). The envelope of the interferogram results from dispersion of the nonlinear crystal. For apodization of the interferogram (steps shown in Fig.\ref{fig:analysis}(b)), a reconstructed interferogram (dark blue) can be calculated by Fourier-transforming the power spectral density of the reference spectrum (Fig. \ref{fig:spectra}).}\label{fig:interferogram}
\end{figure}

\begin{figure}[t]
	\centering
	\fbox{\includegraphics[scale=0.9]{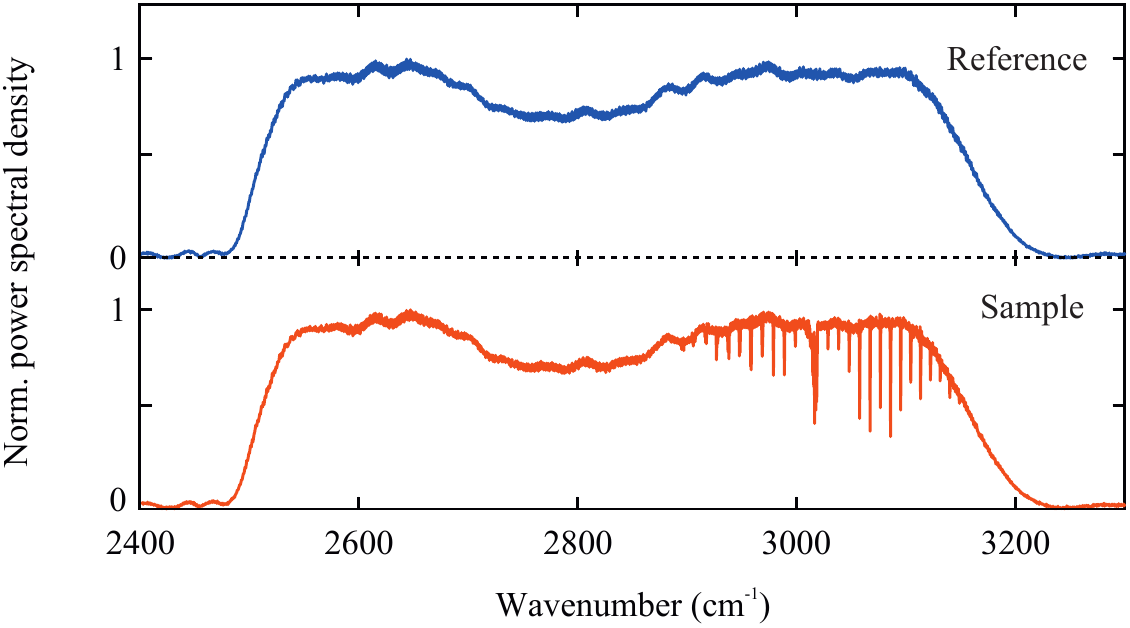}} 
	\caption{Fourier-transformed spectra: Normalized power spectral density calculated by a Fourier-transform of the interferograms measured with pure nitrogen (reference, blue curve) and a mixture of 1\,\% methane in nitrogen (sample, orange curve). Both interferograms were averaged over 100 measurement scans.}\label{fig:spectra}
\end{figure}

Figure \ref{fig:analysis} gives an overview over the steps taken in the measurement and analysis procedure, which will be described in the following.
\paragraph{Measurements and interferogram analysis.}
	As a first step, a number $n$ of interferograms are recorded with a nitrogen-filled sample cell (reference measurement). In an acquisition time of \SI{9}{\s}, the idler mirror is moved with a constant velocity of \SI{2}{\mm/\s}. The available displacement range of the double-sided interferogram results to \SI{18}{\mm}, corresponding to a theoretical maximum resolution of \SI{0.56}{\wn}.
	Each time-referenced interferogram is then transformed into a position-referenced interferogram by using the zero-crossings of the pump interference. In order to increase the signal-to-noise ratio (SNR), the interferograms are added coherently. As an exemplary case, we will conduct the following analysis steps with 100 interferogram scans.
	Since the position reference of each interferogram yields only a relative position information and the absolute position can be shifted by an unknown offset, the correlation of each interferogram is calculated.
	The interferograms are then shifted by the displacement offset value which maximizes the correlation and added.\\
	A normalized section of the resulting interferogram is shown in Fig. \ref{fig:interferogram} as the light blue curve.
	While a classical FTIR interferogram shows a clear and symmetric centerburst, the interferogram presented here is broadened and has an asymmetric shape. This is caused by the dispersion of the nonlinear crystal which is inherent to the system and results in a wavelength(-pair) dependent optical path difference between signal and idler beam throughout the large emission bandwidth. 
	For a 10-$\ur{mm}$-long MgO-doped lithium niobate crystal, the optical path difference between signal and idler beam (both extraordinary polarized) can be calculated to \SI{0.64}{\mm} at \SI{3200}{\wn} idler frequency (about \SI{3.1}{\um} wavelength) and \SI{1.03}{\mm} at \SI{2500}{\wn} idler frequency (\SI{4}{\um} wavelength) using refractive index equations from Ref.\cite{Gayer.2008}. The offset of the points with zero path difference between signal and idler results to \SI{390}{\um}, which is in agreement to the observed width of the interferogram. A chirped interferogram yields the same spectral information as an interferogram without dispersion, 
	which has also been demonstrated for classical FTIR spectroscopy \cite{Griffiths.2007,Sheahen.1975}.
\paragraph{Fourier transform and non-apodized transmission spectrum.} 
	In the next step, the position-referenced and averaged interferogram is Fourier-transformed. We use a discrete Fourier-transform (DFT) based on the Fast Fourier-Transform (FFT) algorithm to obtain the spectrum in spatial frequency space. 
	The power spectral density (PSD) is calculated as the modulus of the complex amplitude resulting from the Fourier-transform. The normalized power spectral density of the reference spectrum (calculated from the interferogram shown in Fig. \ref{fig:interferogram}) is shown in Fig. \ref{fig:spectra} (blue curve).
	The spectrum exhibits a large bandwidth of over \SI{725}{\wn}, centered around \SI{2850}{\wn}.\\
	The measurement and analysis procedure is then repeated with the gas sample filled into the sample cell.
	For a demonstration of a realistic measurement task using high spectral resolution, we chose a sample of 1\,\% methane in nitrogen at atmospheric pressure. The normalized power spectral density of the sample spectrum, measured with 100 scans, is depicted in Fig. \ref{fig:spectra} as the orange curve. The spectrum shows the same envelope and spectral width as the reference spectrum. In the spectral range around \SI{3000}{\wn}, absorption lines are clearly visible.\\
\begin{figure}[tb]
	\centering
	\fbox{\includegraphics[scale=0.9]{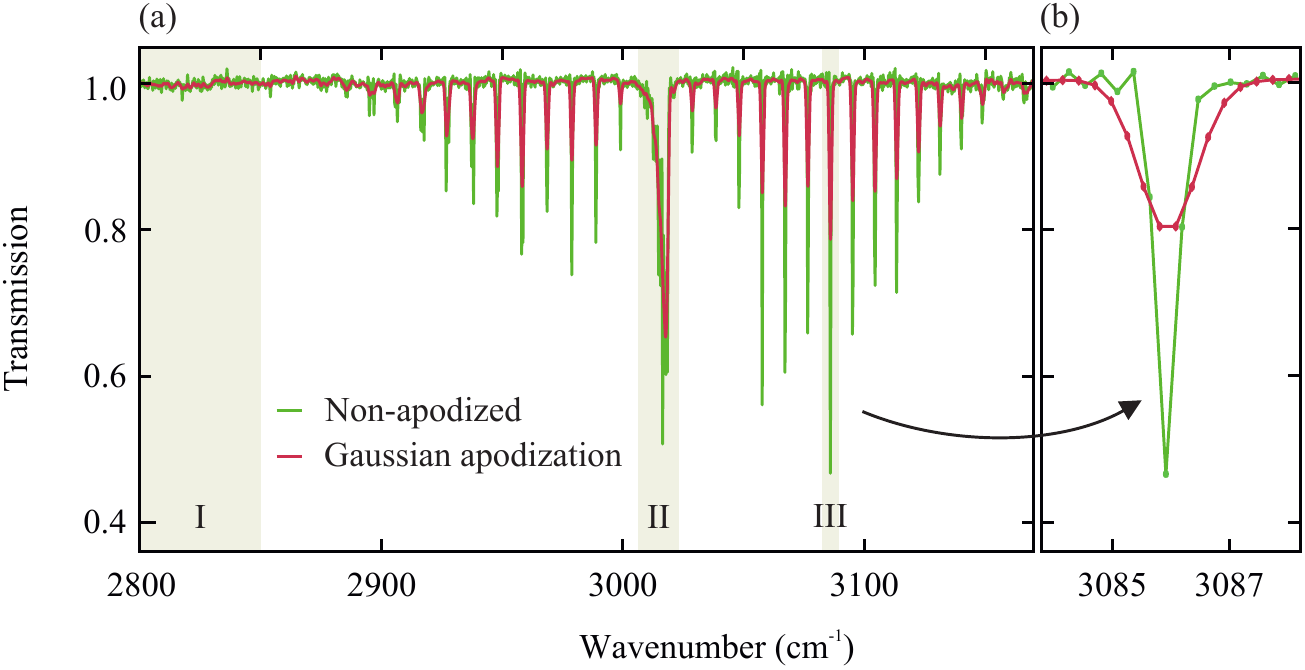}} 
	\caption{Methane transmission spectrum calculated from the spectra shown in Fig. \ref{fig:spectra} without apodization (green) and using a Gaussian apodization function (red). The spectral ranges I-III highlighted in part (a) are further evaluated in section \ref{sec:char}. Part (b) shows a detailed view of the absorption feature in spectral range III.}
	\label{fig:transm_ov}
\end{figure}%
	The transmission of the sample can be calculated from the quotient of the power spectral densities of the reference and sample spectrum. Since the corresponding interferograms were not multiplied with any apodization function, they are truncated by a boxcar (rectangular) window function given by the measurement range. Figure \ref{fig:transm_ov} shows the non-apodized transmission spectrum (green curve) containing rotational absorption lines of the $\nu_3$ band of methane calculated from the spectra shown in Fig. \ref{fig:spectra}. 
\paragraph{Apodization.}
	If the spectral resolution is lower than the width of the measured spectral features, spectra using a simple rectangular window function show pronounced sidelobes. In practical FTIR spectroscopy, these are often removed using apodization (multiplying the interferogram with a function which is unity at zero path difference and decreases with increasing delay), at the cost of a lower spectral resolution depending on the used apodization function \cite{Griffiths.2007}. Direct apodization of a chirped interferogram poses difficulties, as the apodization function would need to be a function of both position and frequency \cite{Griffiths.2007,Sheahen.1975}. 
	Instead, we use a different procedure for calculating the apodized transmission spectrum, which is shown schematically in Fig. \ref{fig:analysis}(b). First, we reconstruct the interferograms without chirp from the spectra shown in Fig. \ref{fig:spectra}. By calculating the power spectral density (or modulus) of the complex amplitude of the spectrum, the phase information is erased. Therefore, a Fourier-transform of the spectrum (into the spatial domain) contains the same spectral information without any phase shift, in essence removing the chirp (cf. \cite{Sheahen.1975}). The resulting reconstructed normalized interferogram of the reference measurement is shown in Fig. \ref{fig:interferogram} (dark blue curve). The width of the interferogram is only determined by the large bandwidth of the SPDC source.\\
	As a next step, the interferograms are then multiplied with an apodization function. For this demonstration we chose a Gaussian apodization function described by
	\begin{equation} \label{eq:apo}
	a(x) = \exp \left(-\frac{(x-L/2)^2}{2(\alpha L)^2}\right),
	\end{equation}
	for the mirror position $x$, the total interferogram length $L=$ \SI{18}{\mm} and parameter $\alpha = 0.2$. The apodized interferograms are then Fourier-transformed. Figure \ref{fig:transm_ov} shows the transmission calculated from the sample and reference spectrum using apodization (red curve), the right part of the graphic provides a detailed view on a single absorption feature. In comparison to the transmission spectrum without apodization (green curve), the spectrum shows no additional sidelobes and a smoother shape.	Due to the decreased spectral resolution, the apodized transmission spectrum shows weaker absorption lines and less noise.

\section{System characterization}\label{sec:char}
	In the following, we use the methane transmission spectra for a characterization of the performance and limitations of the spectrometer. Since the signal-to-noise ratio and spectral resolution depend on the instrument function of the spectra (determined by the window function applied to the interferograms), we use the transmission spectrum without apodization for the analysis of these parameters.
	
\subsection{Signal-to-noise ratio}\label{sec:SNR}
An important characteristic of the nonlinear spectrometer is the signal-to-noise ratio (SNR), which determines the minimum transmission change that can be distinguished from random fluctuations. The signal-to-noise ratio can be quantified from the measured spectra in a spectral range without sample absorption. In the following, we limit the SNR-calculation to a spectral range of \SIrange{2800}{2850}{\wn}, which is highlighted in Fig. \ref{fig:transm_ov} (I). In this range, the measured transmission values (green curve) are normally distributed around their expectation value $T_0\approx1$. The standard deviation $\sigma$ and mean $\mu$ of the measured transmission values allow a calculation of the SNR:
\begin{equation}
\ur{SNR} = \frac{\mu}{\sigma}.
\end{equation}
For the non-apodized transmission spectrum shown in Fig. \ref{fig:transm_ov} (green), the SNR results to 180. This corresponds to a transmission change of about $1.6\,\%$ ($3\sigma$ width) that can be distinguished from noise. The signal-to-noise ratio increases with the square root number of averaged scans.\\
For light sources with low output power, shot noise due to random fluctuations of the photon number $N$ can become a fundamental limitation. In the following, we will derive an estimation for the SNR using parameters of the current setup, assuming shot noise as the only source of noise.
Since the Poisson-distributed number of Photons $N$ has a standard deviation of $\sqrt{N}$, a noise-equivalent power of $P_{\ur{N}}$ can be defined as
\begin{equation}
P_{\ur{N}} = \sqrt{\frac{\epsilon Ph\nu}{t}}\sqrt{F},
\end{equation} 
using the detector quantum efficiency $\epsilon$, the SPDC power $P$, the median energy of a single photon $h\nu$ and the measurement time $t$. The excess noise factor $\sqrt{F}$ describes the increase in shot noise due to the avalanche effect of the photo detector \cite{HamamatsuPhotonics.2004,Kocak.2016}. For a silicon APD, the factor is typically specified with $F\approx 5$ \cite{PerkinElmer.2003}.\\ 
The signal power available for a Fourier-transform measurement with a spectral resolution of $\Delta\vnu$ can be described by (cf. Ref. \cite{Griffiths.2007})
\begin{equation}
S = \eta \epsilon U_{\vnu} \Delta\vnu,
\end{equation}
with the interferometer efficiency $\eta$ and the spectral power density $U_{\vnu}$. The spectral power density of the SPDC source is approximated by
\begin{equation}
U_{\vnu} = \frac{P}{\vnu_{\ur{max}}-\vnu_{\ur{min}}},
\end{equation}
with the SPDC bandwidth $\vnu_{\ur{max}}-\vnu_{\ur{min}}$, which assumes a uniform distribution. Using these approximations, we can estimate a limitation to the SNR for $n$ averaged measurements:
\begin{equation}
\ur{SNR}_{\ur{th}} = \sqrt{n} \frac{S}{P_{\ur{N}}} = \eta \sqrt{\frac{nt}{F}}\sqrt{\frac{\epsilon P}{h\nu}} \frac{\Delta \vnu}{\vnu_{\ur{max}} - \vnu_{\ur{min}}}.
\end{equation}
Estimating the efficiency by the measured maximum interference contrast of $\eta\approx0.15$, for 100 averaged scans with a measurement time of \SI{9}{\s} per scan, a detector efficiency of $\epsilon = 55.8\,\%$, a total SPDC power of \SI{20}{\nano\W}, a single photon energy of \SI{1.24}{\electronvolt} (signal wavelength approximated to \SI{1}{\um}), a spectral resolution of \SI{0.56}{\wn} and a bandwidth of \SI{725}{\wn}, the resulting SNR-limit amounts to $\ur{SNR}_{\ur{th}}\approx 366$. In addition to shot noise limitations, other noise sources such as amplitude noise of the pump laser, detector and electronics dark noise contribute to a lower signal-to-noise ratio. Taking into account the uncertainties in the above derivation, shot noise must be considered an essential contribution to noise in the current setup, in contrast to classical FTIR spectrometers, which are typically limited by detector dark noise \cite{Griffiths.2007}.

\subsection{Spectral resolution}\label{sec:res}
\begin{figure}[t]
	\centering
	\fbox{\includegraphics[scale=0.9]{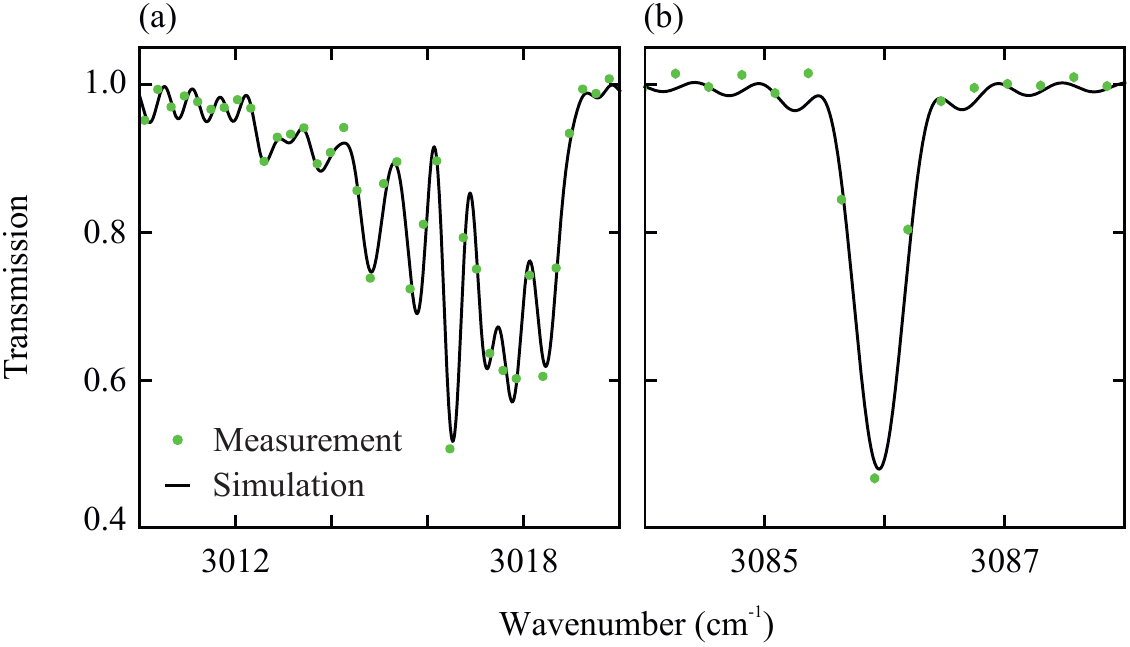}} 
	\caption{Detailed view of the measured transmission spectrum without apodization (green dots) and fitted model function (black curve). Parts (a) and (b) show the spectral ranges II and III highlighted in Fig. \ref{fig:transm_ov}(a). The model function is based on a convolution of spectroscopic data and the sinc-shaped instrument function (Eqs. \ref{eq:fit},\ref{eq:ILS}) and allows determining the maximum spectral resolution (\SI{0.56}{\wn}) of the measured data.}
	\label{fig:res}
\end{figure}

For a quantitative characterization of the spectral resolution, we compare the measured transmission spectrum to the expected transmission spectrum $T_{\ur{m}}$ calculated as a convolution of spectroscopic data $T_{\ur{th}}$ and the instrument function $f$:
\begin{equation}\label{eq:fit}
T_{\ur{m}}(\vnu) = (T_{\ur{th}} * f)(\vnu).
\end{equation}  
Spectroscopic data on methane $T_{\ur{th}}(\vnu)$ is obtained from the HITRAN database \cite{Hitran, HitranOTW}, using a Voigt line profile and ambient conditions as parameters (room temperature of \SI{24}{\degreeCelsius}, 1\,atm pressure). Under these conditions, the widths of individual absorption lines is at the order of \SI{0.12}{\wn} (full width at half maximum, FWHM). As described before, we will analyze the transmission spectrum which was calculated using no apodization, which yields the highest spectral resolution. The instrument function $f(\vnu)$, which is defined as the Fourier-transform of the rectangular window function, can then be described by \cite{Griffiths.2007}
\begin{equation}\label{eq:ILS}
f(\vnu) = \frac{2}{\Delta \vnu}\cdot \text{sinc}\left(\frac{2\pi\vnu}{\Delta\vnu}\right),
\end{equation}
with the spectral resolution $\Delta\vnu$, which is limited to the inverse of the maximum interferometer delay $1/L\approx $ \SI{0.56}{\wn}.\\
In order to determine the spectral resolution of the measured transmission spectrum, the model function (Eq. \ref{eq:fit}) is fitted to the measurement data using a non-linear least square optimization. The optimization yields a spectral resolution of $\Delta \vnu =$ \SI{0.56}{\wn}, which is in agreement to the theoretic value. This shows that the spectral resolution of the nonlinear interferometer is only limited by the maximum delay between the interferometer arms, in contrast to previous implementations of nonlinear interferometers using grating spectrometers with lower spectral resolution \cite{Kalashnikov.2016b, Paterova.2018b}.\\
Fig. \ref{fig:res} shows the measured transmission spectrum (green dots) and the fitted model function (black curve) for the spectral ranges II and III marked in Fig. \ref{fig:transm_ov}.
Clearly visible are the sidelobes around the absorption feature in Fig. \ref{fig:res}(b) due to the shape of the instrument function. 
A systematic influence of methane absorption features on the difference between measured transmission and fitted model function is still observed, which we attribute to small deviations in the shape of the instrument function.

\subsection{Accuracy}\label{sec:acc}
\begin{figure}[t]
	\centering
	\fbox{\includegraphics[scale=0.9]{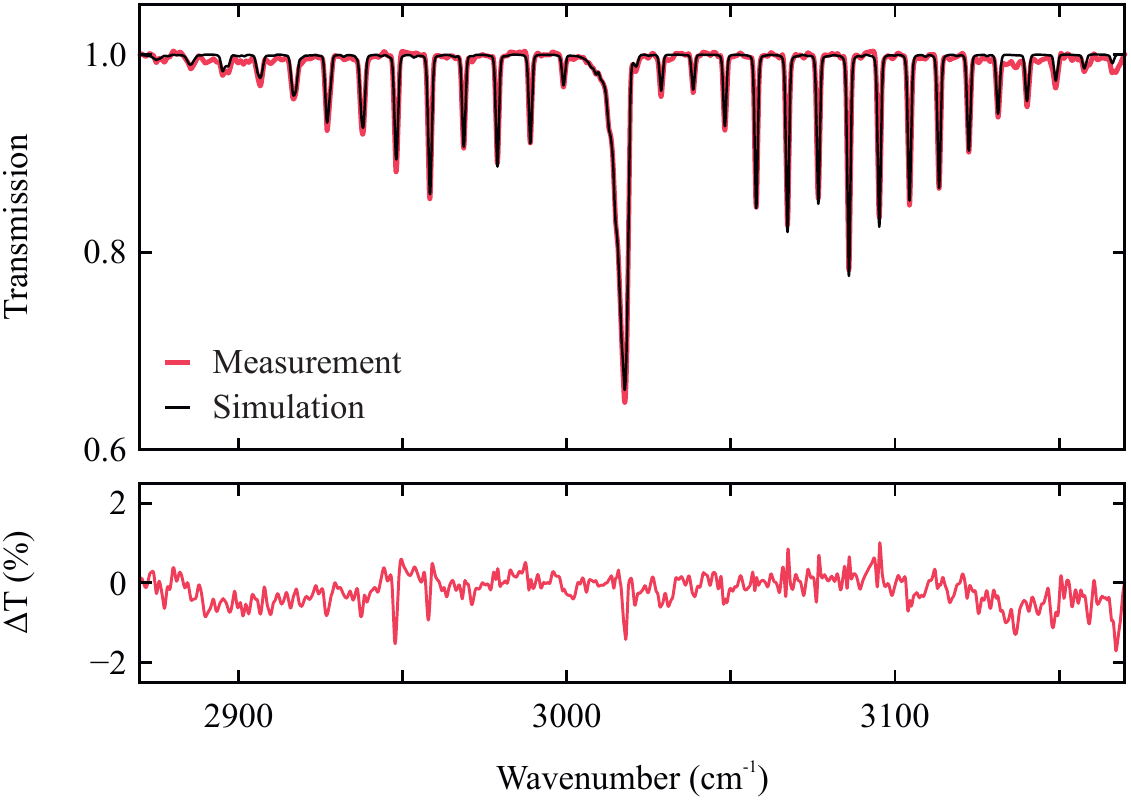}} 
	\caption{Measured transmission spectrum of methane using a Gaussian apodization function (red curve) which reduces the spectral resolution to about \SI{1}{\wn} (full width at half maximum). The black curve shows the model transmission function, based on a convolution of spectroscopic data and the Gaussian instrument function (Eq. \ref{eq:ils_gauss}), which was fitted to the measured transmission data. The transmission residuum $\Delta T$ is shown in the lower graph (red curve).}
	\label{fig:transm}
\end{figure}
In order to determine the transmission accuracy of the nonlinear interferometer, we use the apodized transmission spectrum (shown as red curves in Figs. \ref{fig:transm_ov},\ref{fig:transm}), for which the amplitude of sidelobes is greatly reduced.
For a quantitative evaluation, analogue to the procedure described above, we calculate the expected spectrum from a convolution (Eq. \ref{eq:fit}) of spectroscopic data \cite{Hitran,HitranOTW} and a modified instrument function. The instrument function of the apodized interferogram is the Fourier-transform of the apodization function (Eq. \ref{eq:apo}), which can be approximated by a Gaussian function:
\begin{equation}\label{eq:ils_gauss}
f(\vnu) \approx \frac{1}{\sqrt{2\pi}\sigma_{\ur{f}}} \exp\left(-\frac{\vnu^2}{2\sigma_{\ur{f}}^2}\right),
\end{equation}
with an expected width of $\sigma_{\ur{f}} = 1/(2\pi\alpha L)\approx0.44\,\ur{cm}^{-1}$. The model function is fitted to the measured data using the width of the gaussian instrument function as free parameter. The optimization yields $\sigma_{\ur{f}} =$ \SI{0.43}{\wn}, which is in good agreement to the expected value.\\
The resolution of the spectrum (full width at half maximum of the gaussian instrument function) results to $2\sqrt{2\ln(2)}\sigma_{\ur{f}}\approx$ \SI{1}{\wn}. Analogue to the procedure described in section \ref{sec:SNR}, we can determine the signal-to-noise ratio from the transmission spectrum in a spectral range of \SIrange{2800}{2850}{\wn}. The SNR of the apodized transmission spectrum results to $390$, which corresponds to a transmission change of \SI{0.8}{\percent} (3$\sigma$ width) which can be distinguished from noise.\\
The apodized transmission spectrum (red curve) and the model function using the optimized parameter (black curve) are shown in Fig. \ref{fig:transm}.
The transmission residuum (difference between simulation and measured values) is shown below the transmission spectrum. Over a broad spectral range, the residuum shows small deviations distributed around the baseline. For few absorption lines in the P- and Q-branch, systematic deviations between measured data and model function can be observed. The standard deviation of the residuum amounts to $0.34\,\%$, from which we can estimate the accuracy of the transmission values to $1\,\%$ (3$\sigma$ width). The transmission spectrum is slightly less accurate than expected from noise contributions alone, due to few outliers. The overall good accuracy, in combination with its high spectral resolution and large spectral bandwidth, demonstrates the applicability of the nonlinear interferometer to quantitative spectroscopic tasks.

\section{Conclusion}
Nonlinear interferometers using correlated photon pairs far from degeneracy offer new possibilities for mid-infrared metrology. Here, we have demonstrated Fourier-transform mid-infrared transmission spectroscopy of a gaseous sample using near-infrared single-pixel detection. 
Our approach combines the data acquisition and analysis in analogy to classical FTIR-spectroscopy with a quantum effect: The phase and transmission dependence of the two-photon interference allows measuring mid-infrared information with near-infrared detection.\\ 
For a performance characterization, the transmission spectrum of methane around \SI{3.3}{\mum} was measured with its rotational lines resolved.
The current proof-of-concept implementation provides a spectral coverage of \SIrange{3.1}{4.0}{\um} (corresponding to an instantaneous bandwidth of over \SI{725}{\wn}) and a spectral resolution better than \SI{1}{\wn}. The transmission spectrum using a Gaussian apodization function shows good agreement to a simulated spectrum based on HITRAN data. The residuum level below $1\,\%$ (3$\sigma$ width) demonstrates the photometric accuracy over a broadband spectral range and the absence of nonlinearities in the transmission measurement.\\
The spectral resolution of the nonlinear interferometer is only limited by the available maximum delay between the interferometer arms. While the demonstrated sub-wavenumber resolution allows for practical gas analysis in its current state, the resolution can be further improved with a setup allowing for larger mirror displacement. Here, the Fourier-transform approach shows a clear advantage over grating-based solutions as used in other implementations of nonlinear interferometers \cite{Kalashnikov.2016b,Paterova.2018b}, but also compared to upconversion spectrometers \cite{Tidemand.2016,Friis.2019,Wolf.2017}.\\
Due to the low output power of the SPDC source, photon shot noise is demonstrated to be a significant noise contribution. Classical FTIR spectrometers as well as upconversion spectrometers offer considerably higher signal-to-noise ratios.
Future implementations of nonlinear interferometers may greatly benefit from more efficient sources for correlated photons, which could be realized using other nonlinear materials or waveguides \cite{Leverrier.2006}.
Using a high-gain nonlinear interferometer (cf. \cite{Machado.2020}) or resonant pump enhancement may also yield increased measurement signal.\\
In the future, nonlinear interferometers may prove a viable alternative for spectroscopic applications in specialized spectral ranges, if the signal-to-noise ratio can be increased.
The infrared power used to illuminate the sample in the presented setup is six orders of magnitude lower compared to a classical FTIR spectrometer, making it suitable for the analysis of sensitive samples. 
Another interesting perspective is the application of the demonstrated Fourier-transform approach to hyperspectral wide-field microscopy with correlated photons (cf. \cite{Paterova.2020,Kviatkosvky.2020}).\\
With respect to future applications, the interferometer scheme may be realized more cost-efficiently by replacing the Ti:sapphire pump laser with a single-frequency diode laser reaching output powers of several hundreds of milliwatts, since tunability of the pump laser is not needed. Using a different pump wavelength and poling period of the nonlinear crystal, the full infrared transparency range of lithium niobate can be addressed.
The measurement approach can be extended to other spectral ranges, such as the terahertz regime \cite{Kutas.2020, Kutas.2020b}, or into the fingerprint infrared range from \SIrange{7}{12}{\um} wavelength using other nonlinear materials.\\
The presented measurement approach enables the novel measurement principle of nonlinear interferometers, allowing mid-infrared measurement via near-infrared detection, to be applied to spectroscopic tasks efficiently and accurately.

\section*{Funding}
Fraunhofer-Gesellschaft (Lighthouse project QUILT).


\section*{Disclosures}
The authors declare no conflicts of interest.

\bibliography{literature}

\end{document}